%% file: TTbChiralBosons.tex
\documentclass[a4paper,11pt]{article}
\usepackage[utf8]{inputenc}
\pdfoutput=1

\usepackage{jheppub,float} 
\usepackage{tikzit}
\input{flowchart.tikzstyles}
\usepackage{xcolor}

\usepackage{parskip}

\usepackage{tocloft}

\usepackage{hyperref}
\usepackage{slashed}
\usepackage{mathrsfs}

\newcommand\numberthis{\addtocounter{equation}{1}\tag{\theequation}}

\usepackage{cleveref}

\usepackage{tikz}
\usetikzlibrary{arrows,shapes}
\usetikzlibrary{positioning,calc}

\renewcommand{\a}{\alpha}
\renewcommand{\b}{\beta}
\newcommand{\s}{\sigma}

\newcommand{\e}{\epsilon}

\newcommand{\vt}{\vartheta}

\newcommand{\mbb}[1]{\mathbb{#1}}
\newcommand{\mbf}[1]{\mathbf{#1}}
\newcommand{\mc}[1]{\mathcal{#1}}
\newcommand{\ttb}{\text{T}\overline{\text{T}}}
\newcommand{\dd}{\text{d}}

\title{Irrelevant Deformations of Chiral Bosons}

\author[a]{Subhroneel Chakrabarti,}
\author[b]{Divyanshu Gupta,}
\author[a]{Arkajyoti Manna,}
\author[c]{\\and Madhusudhan Raman}
\affiliation[a]{Institute of Mathematical Sciences, Homi Bhabha National Institute (HBNI)\\
	IV Cross Road, C.~I.~T.~Campus, Taramani, Chennai 600 113, Tamil Nadu, India}
\affiliation[b]{BITS-Pilani, KK Birla Goa Campus\\
	NH 17B, Bypass Road, Zuarinagar 403 726, Goa, India}
\affiliation[c]{Department of Theoretical Physics, Tata Institute of Fundamental Research\\
	Homi Bhabha Road, Navy Nagar, Colaba, Mumbai 400 005, Maharashtra, India\\}

\emailAdd{subhroneelc@imsc.res.in}
\emailAdd{guptadv17@gmail.com}
\emailAdd{arkajyotim@imsc.res.in}
\emailAdd{madhur@theory.tifr.res.in}

\abstract{We study $ \ttb $ deformations of chiral bosons using the formalism due to Sen. For arbitrary numbers of left- and right-chiral bosons, we find that the $ \ttb $-deformed Lagrangian can be computed in closed form, giving rise to a novel non-local action in Sen's formalism. We establish that at the limit of infinite $\ttb$ coupling, the equations of motion of deformed theory exhibits chiral decoupling. We then turn to a discussion of $\ttb$-deformed chiral fermions, and point out that the stress tensor of the $ \ttb $-deformed free fermion coincides with the undeformed seed theory. We explain this behaviour of the stress tensor by noting that the deformation term in the action is purely topological in nature and closely resembles the fermionic Wess-Zumino term in the Green-Schwarz formalism. In turn, this observation also explains a puzzle in the literature, viz.~why the $\ttb$ deformation of multiple free fermions truncate at linear order. We conclude by discussing the possibility of an interplay between $ \ttb $ deformations and bosonisation.}

\preprint{\texttt{TIFR/TH/20-46}}

\begin{document} 
\maketitle
\flushbottom

\section{Introduction}

The subject of soluble irrelevant deformations of two-dimensional quantum field theories has, in recent year, enjoyed significant attention. Various aspects of the so-called $ \ttb $ deformation --- triggered at each step of the ``flow'' by the determinant of the stress tensor \cite{Zamolodchikov:2004ce,Cavaglia:2016oda} --- such as the finite-volume spectrum \cite{Smirnov:2016lqw} and the S-matrix of the deformed theory \cite{Dubovsky:2012wk,Dubovsky:2013ira,Dubovsky:2017cnj} are known exactly. These deformations have been studied from a various points of view, for example holography \cite{McGough:2016lol,Kraus:2018xrn,Taylor:2018xcy,Hartman:2018tkw,Caputa:2019pam}, supersymmetric quantum field theories \cite{Baggio:2018rpv,Chang:2018dge,Jiang:2019hux,Chang:2019kiu,Ferko:2019oyv,Coleman:2019dvf}, and string theory \cite{Callebaut:2019omt,Blair:2020ops}. An excellent introduction to this field can be found in \cite{Jiang:2019hxb}.

The goal of this paper is to study $ \ttb $ deformations of chiral bosons. In this introduction, we will motivate our analyses, situating our own study in the context of recent developments in this and adjacent fields.

\subsection*{Chiral Decoupling}
In \cite{Chakrabarti:2020pxr}, we explored a curious aspect of the $ \ttb $ deformation: the limit in which the $ \ttb $ coupling constant $ \lambda $ is sent to infinity. In order to make sense of the above limit, it was imperative that we studied the deformed theory within the Hamiltonian formalism. We argued that this limit generically effects a decoupling of left- and right-chiral degrees of freedom. Schematically, we may represent our findings as follows:
\begin{center}
	\begin{tikzpicture}[scale=0.19]
	\tikzstyle{every node}+=[inner sep=0pt]
	\node[text width=3cm, align = center] at (-15.6,-19.2) {  Free Bosons}; 
	\draw [black] (-8.9,-19.2) -- (-2.8,-19.2); 
	\fill [black] (-2.0,-19.2) -- (-2.8,-18.7) -- (-2.8,-19.7); 
	\draw (-5.5,-18.7) node [above] {$\mathrm{T}\overline{\mathrm{T}}$};
	\node[text width=5cm, align = center] at (8.3,-19.2) { Nambu-Goto String}; 
	\node[text width=3cm, align = center] at (34.3,-19.2) { Chiral Bosons}; 
	\draw [black] (26.3,-19.2) -- (18.6,-19.2);
	\fill [black] (26.3,-19.2) -- (25.5,-18.7) -- (25.5,-19.7);
	\draw (22.00,-18.7) node [above] {$\lambda \rightarrow \infty$};
	\end{tikzpicture} 
\end{center}
A similar progression was observed for the case of free fermions:
\begin{center}
\begin{tikzpicture}[scale=0.19]
\tikzstyle{every node}+=[inner sep=0pt]
\node[text width=3cm, align = center] at (-15.6,-19.2) {Free Fermions}; 
\draw [black] (-8.9,-19.2) -- (-2.8,-19.2); 
\fill [black] (-2.0,-19.2) -- (-2.8,-18.7) -- (-2.8,-19.7); 
\draw (-5.5,-18.7) node [above] {$\mathrm{T}\overline{\mathrm{T}}$};
\node[text width=5cm, align = center] at (8.3,-19.2) {Interacting Fermions}; 
\node[text width=3cm, align = center] at (34.3,-19.2) {Free Fermions}; 
\draw [black] (26.3,-19.2) -- (18.6,-19.2);
\fill [black] (26.3,-19.2) -- (25.5,-18.7) -- (25.5,-19.7);
\draw (22.00,-18.7) node [above] {$\lambda \rightarrow \infty$};
\end{tikzpicture} 
\end{center}
The curious nature of these progressions, from free to interacting to free theory, naturally motivates us to ask: what next? That is, what happens to a theory of free chiral bosons under a $ \ttb $ deformation? 

We will argue in this section that the choice to focus on this question places us at the intersection of various interesting aspects of the $ \ttb $ deformation that merit further investigation.

\subsection*{A Choice of Formalism}
Now that we are decided on the study of chiral bosons, we must first decide how to describe them. In the study of chiral fields, it is well-known that the constraint of chirality precludes the possibility of their study via the usual methods of free field theory. This is true more broadly of all self-dual $ (2n+1)$-form field strengths in $ 4n+2 $ spacetime dimensions. (The case $ n = 0 $ is a theory of chiral bosons.) 

This subject of self-dual field strengths has an interesting and tortured history, with theorists caught for decades between Scylla and Charybdis. The earliest investigations \cite{Siegel:1983es,Henneaux:1988gg,Schwarz:1993vs} used actions that were not manifestly Lorentz invariant. More recently, too, \cite{Ouyang:2020rpq} studied $ \ttb $ deformations of chiral bosons utilizing the formalism of Floreanini and Jackiw \cite{Floreanini:1987as,Sonnenschein:1988ug}.

The lack of manifest Lorentz invariance was subsequently overcome \cite{McClain:1990sx,Pasti:1996vs} at the cost of the introduction of auxiliary fields: either an infinite number of them, or a finite number of auxiliary field with non-polynomial actions. For the better part of the last two decades, this was the most widely accepted framework for the discussion of chiral $ p $-form fields despite the lack of off-shell self-duality; this permitted a classical analysis but complicates the discussion of the corresponding quantum theory. Another recent proposal for a Lorentz invariant action for chiral bosons was put forward in \cite{Townsend:2019koy}.

Following the first successful construction of the quantum Batalin–Vilkovisky master action for closed superstring field theory \cite{Sen:2015uaa}, the aforementioned difficulties were overcome in the context of Type IIB SUGRA by Sen in \cite{Sen:2015nph}. The construction was readily extended to theory of self-dual forms with arbitrary rank in \cite{Sen:2019qit}. Sen's formalism, in addition to the $ (2n+1) $-form field strength, also contains an additional $ (2n) $-form field with the wrong sign for the kinetic term. Despite this, it was shown that the (free) dynamics of this additional field completely decouples --- even from the background metric, and in the quantum theory as well --- thereby forming an autonomous sector inert to the dynamics of other physical interacting fields.

We have elected to use Sen's formalism in our study of chiral bosons. This choice was made in part by the demands of the problem at hand: various aspects of $ \ttb $ deformations are quantum mechanically exact, and it is desirable to work with a formalism that is capable, at least in principle, of eventually accessing the quantum theory. Chirality is baked into the theory, i.e.~the field variables are self-dual off-shell, and not imposed as an equation of motion. Additionally, manifest Lorentz invariance and polynomial nature of the action makes the case for adoption of Sen's formalism very strong indeed.\footnote{Sen's formalism has been successfully used in the context of self-dual $3$-forms in six-dimensional $(2,0)$ theories, see \cite{Lambert:2019diy,Andriolo:2020ykk,Gustavsson:2020ugb}.}

To work with a formalism where the self-duality of the chiral boson is imposed at the level of an equation of motion (i.e.~on-shell) is to foreclose on the possibility of studying the quantum aspects of this theory within this formalism. With this perspective, while the demonstration of \cite{Ouyang:2020rpq} is elegant, it is not surprising, since at the level of the classical Lagrangian (an off-shell quantity), we are dealing with a decomposition of the scalar field that is only permissible on-shell. This is evident in the relation between the Hamiltonian densities of the deformed free scalar on one hand and the deformed left-chiral and right-chiral scalar on the other, see \cite[eq.~(20)]{Ouyang:2020rpq}. 

Another way of seeing this is that for \cite{Ouyang:2020rpq}, the seed theory described by the Floreanini-Jackiw Lagrangian is a theory of chiral bosons \emph{on-shell}, and that this is no longer the case once the $ \ttb $ flow begins. This is because in the Floreanini-Jackiw formalism, chirality is imposed on-shell by solving an equation of motion. Once deformations are turned on, the equation of motion that imposes chirality changes.

Our set-up is quite different. While the on-shell dynamics of the formalisms of Floreanini-Jackiw and Sen agree, in the latter formalism one works with fields that are self-dual \emph{off-shell}, i.e.~they continue to be off-shell once deformations are turned on. This off-shell self-duality imposes severe constraints on the form of the interactions one can introduce in Sen's formalism. We will see that this implies an interesting closed form of the $ \ttb $-deformed Lagrangian.

\subsection*{Lorentzian vs.~Euclidean Spacetime}
The decision to study chiral bosons in two dimensions \emph{forces} us to work with a Lorentzian signature. It is helpful to quickly review this argument, so let us suppose that we are studying chiral (i.e.~self-dual or anti-self-dual) $ p $-forms on $ \mathbb{M} $, an $ n $-dimensional manifold whose metric has $ s $ time coordinates, and derive this conclusion.

Consider the Hodge star operator $ \star : \Omega^p(\mathbb{M}) \rightarrow \Omega^{n - p}(\mathbb{M}) $, which maps $ p $-forms to $ (n-p) $-forms. We find that
\begin{equation}\label{key}
	\star^2 = (-1)^{s+p(n-p)} \ ,
\end{equation}
where $ s = 0 $ for Euclidean space and $ s = 1 $ for Lorentzian spacetimes. A self-dual $ (+) $ or anti-self-dual $ (-) $ form $ A $ satisfies the property
\begin{equation}\label{key}
	A = \pm \star A \ ,
\end{equation}
which in turn implies $ n = 2p $, i.e.~we can only have chiral forms in even dimensions. Further, on Hodge dualising the above equation once more, we find that
\begin{equation}\label{key}
	\star^2 = (-1)^{s+p(n-p)} = +1 \ ,
\end{equation}
which in turn implies that for Euclidean signature, $ p $ is even, implying $ n = 4 k $ for $ k \in \mathbb{N} $. However, for the Minkowski signature, we find $ p^2 + 1 $ is even, so $ p $ is odd and $ n = 4k+2 $. Therefore, in order to study real chiral bosons in two dimensions, we must work with a Lorentzian signature.

In contrast, the $ \ttb $ deformation has been most rigorously studied in Euclidean space. For example, one typically defines the composite $ \ttb $ operator via point-splitting and the proof of Zamolodchikov's factorization formula \cite{Zamolodchikov:2004ce} uses Euclidean operator product expansions. In the Lorentzian setting, it is natural to expect additional lightcone singularities. To the best of our knowledge, such an analysis has not yet been carried out.

Despite this, various Lorentzian aspects of the $ \ttb $ deformation --- defined in the same way, except on Minkowski spacetime --- have been studied in the literature. As illustrative examples:  \cite{McGough:2016lol} builds on \cite{Cardy:2015xaa} and studies signal propagation speeds; \cite{Frolov:2019nrr} interprets the inviscid Burgers equation which determines the spectrum of the $ \ttb $-deformed theory as encoding the gauge invariance of the target spacetime energy and momentum of a non-critical string theory quantised in a one-parameter family of deformations of the standard lightcone gauge called the uniform lightcone gauge; finally, \cite{Kruthoff:2020hsi} adopt a Hamiltonian point of view on $ \ttb $ deformations and show that their results in Lorentzian spacetime are consistent with known exact results like the Castillejo-Dalitz-Dyson phase that relates the deformed and undeformed S-matrices.

Our perspective in this note will be similar.

\subsection*{$\ttb$ Flow of Chiral Bosons, Fermions, and Bosonisation}
Another interesting aspect of Sen's formalism is that the classical action is \emph{not} of the general form assumed by \cite{Bonelli:2018kik}, i.e.~of the form
\begin{equation}\label{key}
	S = \int \dd^2 x \, \sqrt{g} \, \mathcal{L} \ .
\end{equation}
We will therefore construct the $ \ttb $-deformed Lagrangian order-by-order in the $ \ttb $ coupling constant, with the deformations sourced by the determinant of the stress tensor.

Our central result is that the uniform structure of the deformations implies that the $ \ttb $-deformed Lagrangian for a pair of chiral bosons can in fact be computed in closed form. On considering arbitrary numbers of left- and right-chiral bosons, once again we compute the closed form of the deformed Lagrangian to all orders in the $\ttb$ coupling constant. This result produces a novel non-local action in Sen's formalism, which is intriguing in its own right. We also observe that the $\ttb$-deformed free fermion Lagrangian, which is known to truncate at linear order \cite{Coleman:2019dvf}, does not actually alter the classical stress tensor of the seed theory. To the best of our knowledge, this fact is noted and explicitly demonstrated for the first time in the literature. As we will explain, this feature is due to the fact that the deformation term for free fermion is a purely topological one and therefore incapable of altering the stress tensor.

We then initiate a study of the possible interplay between bosonisation and $\ttb$ deformations. We make a preliminary conjecture that should serve as a suitable starting point for this line of investigation.

Of course, bosonisation duality (say, between sine Gordon theory and the massive Thirring model \cite{Coleman:1974bu}) is a quantum mechanically exact statement about correlation functions. This demonstration is made possible in part by the relevant nature of deformations, which makes the corresponding quantum theories well-defined, i.e.~renormalisable. In this paper, we have restricted our attention only to the classical aspects of $ \ttb $-deformed theories. 

We hope to address quantum mechanical features of this problem, leveraging the analyses of \cite{Aharony:2018vux,Guica:2019vnb,Cardy:2019qao,He:2020udl} on correlation functions and \cite{Rosenhaus:2019utc} on renormalisation, in the near future.

\subsection*{Outline}
We now summarise the contents of our paper. 

In \S\ref{sec:SenFormalismReview}, we provide a pedagogical review of Sen's formalism as it is applied to chiral bosons in two dimensions. Following this, in \S\ref{sec:StressTensor} we explain our strategy for computing the stress tensor. We feel this will help acclimatise our readers to Sen's formalism, which may be unfamiliar. Our main results are contained in \S\ref{sec:TTbChiralBosons}, where we study $ \ttb $ deformations of theories with left- and right-chiral bosons, and solve for the closed form of the deformed Lagrangian. In \S\ref{sec:Bosonisation}, we begin by pointing out that stress tensor of the $ \ttb $-deformed free fermions theory coincides with the undeformed seed theory. We explain that this is due to the fact that the deformation terms in the action is purely topological in nature. We then initiate a study of the possible bosonisation duality relating two $ \ttb $-deformed theories, and speculate on the possibility that bosonisation ``commutes'' with $ \ttb $ deformations. For the theory at hand, this conjecture is not intended to succeed, but to merely provide a stepping stone towards a more concrete realization of this program. We conclude in \S\ref{sec:Discussion} by summarising our results and listing some further directions for research. 

\section{Sen's Formalism for Chiral Bosons}
\label{sec:SenFormalismReview}

A chiral boson in two dimensions is nothing but an $1$-form field $A_g$ which is self-dual, i.e. $A_g = \star_g A_g$. Typically one thinks of this $1$-form as a \emph{field strength} for a scalar, viz.~$A_g \sim \dd\varphi$, where $\dd$ is the exterior derivative and $\varphi$ is a scalar field. However, in Sen's formalism, as we will shortly see,  we need not invoke a rewriting of the self-dual $1$-form as an exact $1$-form. Further, it is sufficient for us to confine our analysis to the classical aspects of this action. Nonetheless, we wish to emphasize that the fully quantum mechanical analysis in the Hamiltonian formulation of this formalism is known as well. For more detailed discussions, we refer the reader to the original papers \cite{Sen:2015nph,Sen:2019qit} and also \cite{Andriolo:2020ykk}.

Sen's action is of the following form
\begin{equation} \label{eq:Sen_action}
	S =  \int_{(\mbb{R}^{(1,1)},\,\eta)} \left[ \dfrac{1}{2}\, \dd \phi \wedge \star_{\eta} \dd \phi  - 2 \, A \wedge \dd\phi + A \wedge \mc{M}(A) \right] \;.
\end{equation}
The action describes a theory defined over a Lorentzian $2$-manifold $\mbb{M}$ equipped with a metric $g_{\mu \nu}$. However, note that the integral is defined over the two-dimensional Minkowski spacetime $\mbb{R}^{(1,1)}$. The field $\phi$ is a scalar field on $\mbb{R}^{(1,1)}$, and the field $A$ is an $1$-form on $\mbb{R}^{(1,1)}$ which is defined to be \emph{self-dual with respect to the flat metric}, i.e.~$A = \star_{\eta} A$. These are just some of the several unfamiliar features of this action. 

Notice that due to this unusual self-duality constraint on the $1$-form $A$, it does not belong to the space of self-dual $ 1 $-forms associated to $(\mbb{M},g)$. Following \cite{Andriolo:2020ykk}, we refer to these objects as \underline{pseudoforms}.

The map $\mc{M}(A)$ is a linear map on the space of self-dual pseudoforms with the following properties (all of which are satisfied at the level of action itself, i.e.~off-shell).
\begin{itemize}
	\item $\mc{M}(A)$ maps self-dual pseudoforms to anti-self-dual pseudoforms. That is, for all $ A = \star_{\eta} A $, we have
	\begin{equation}
		\mc{M}(A) = - \star_{\eta} \mc{M}(A) \;.
	\end{equation}
	\item $\mc{M}$ annihilates anti-self-dual pseudoforms. That is, for all $ B = - \star_{\eta} B $ we have
	\begin{equation}\label{eq:M_on_anti}
	\mc{M}(B) = 0 \;.
	\end{equation}
	\item For any two self-dual pseudoforms $A_1$ and $A_2$, the map satisfies
	\begin{equation}
		A_1 \wedge \mc{M}(A_2) = A_2 \wedge \mc{M}(A_1) \;.
	\end{equation}
	\item Even though $A$ and $\mc{M}(A)$ are pseudoforms, the following specific linear combination belongs to the space of self-dual forms associated to $(\mbb{M},g)$.
	\begin{equation}\label{eq:Ag_selfdual}
		A - \mc{M}(A) = \star_g \left(A - \mc{M}(A)\right) \;.
	\end{equation}
\end{itemize}
The astute reader will conclude from this last property that the map $\mc{M}(A)$ is necessarily cognizant of the background metric, even though the pseudoforms themselves are not. This is indeed true. An explicit construction of the map with the above properties was supplied in \cite{Sen:2015nph} (also see \cite{Sen:2019qit,Andriolo:2020ykk}). We will, however, not require the explicit construction of the map in this paper.

From the action given in \cref{eq:Sen_action}, one can easily derive the equations of motion which are
\begin{align}
	\dd \left[  \dfrac{1}{2} \star_{\eta} \dd\phi + A  \right] &= 0 \;, \\
	\left(1-\star_{\eta}\right) \bigg[ \dd\phi - \mc{M}(A) \bigg] &=0 \;.
\end{align}

Let us define the following fields
\begin{align}
	\hat{A} &:= A + \dfrac{1}{2} \left(\dd\phi + \star_{\eta} \dd\phi\right) \;,\\
	A_g     &:= A - \mc{M}(A) \;. \label{eq:A_g}
\end{align}
In terms of these new fields, the equations of motion assume a very simple form:
\begin{align}
	\dd \hat{A} &= 0 \;,\\
	\dd A_g &= 0 \;. \label{eq:Ag_eom}
\end{align}
Notice that we can rewrite the property in \cref{eq:Ag_selfdual} as $A_g = \star_g A_g$. This is true off-shell. Therefore, we identify the variable $A_g$ as the desired chiral boson on $(\mbb{M},g)$ which satisfies the free massless equation of motion \cref{eq:Ag_eom} as expected.

The field $\hat{A}$ also satisfies the free massless equation of motion and we can see that at the level of equation of motion at least, it is completely decoupled from the physical degree of freedom carried by $A_g$. It was shown in \cite{Sen:2019qit} that this decoupling continues to hold even at the quantum level. Since the only term in the action that is cognizant of the background metric is $\mc{M}(A)$, we see that the field $\hat{A}$ does not couple to the background metric. 

The decoupling of $\hat{A}$ is absolute in this regard; it does not couple to any of the physical fields and is therefore completely inaccessible to any experiment one might wish to perform. For all intents and purposes, the field $\hat{A}$ is just an auxiliary aid which enables us to formulate a description of the dynamics of self-dual form fields which possesses the desirable properties we outlined in the introduction.

There is, however, one minor trade-off we must accept when adopting Sen's formalism: the action is not manifestly invariant under general coordinate transformations. In fact, under diffeomorphisms, the transformation properties of the field $\phi$ and the pseudoform $A$ are rather unusual (the details can be found in \cite{Sen:2015nph,Andriolo:2020ykk}). Despite the lack of manifest general coordinate invariance, one can verify that the action is indeed invariant under general coordinate transformations, as it should be.

Before proceeding, a brief word about our notation and nomenclature is in order. Throughout the rest of the paper the Hodge dual with respect to $\eta$ will be denoted $\star$ and we suppress the $\eta$ subscript for convenience. For the Hodge dual with respect to the background metric $ g $ on $ \mathbb{M} $, we will continue to explicitly write $\star_g$. Furthermore, we suppress the subscripts $\mbb{R}^{(1,1)}$ and $\eta$ on all integrals over Minkowski spacetime. Wherever we use the integral over $\mbb{M}$, we mention it in the subscript or write the diffeomorphism invariant measure factor explicitly. Finally, we will use terms left-chiral (respectively, right-chiral) and self-dual (respectively, anti-self-dual) interchangeably.

\section{The Stress Tensor}
\label{sec:StressTensor}

Despite the unfamiliar structure of the Sen's action, one can compute the associated stress tensor in an arbitrary background following a rather simple prescription. For the case of chiral 3-forms in six dimensions this has already been done in \citep{Andriolo:2020ykk}; it is evident, however, that this analysis can be readily extended to any chiral $p$-form in $2p$ dimensions. For the sake of completeness, we explicitly present the analysis for the case at hand, i.e.~the case of chiral 1-form in two dimensions.

The steps to compute the stress tensor are as follows:
\begin{enumerate}
	\item Consider the standard action for an \emph{arbitrary} 1-form field $\mbf{A}$ living on $\mbb{M}$
	\begin{equation}\label{key}
		S_{\text{aux.}} = \dfrac{1}{2} \int_{\mbb{M},\, g}  \mbf{A} \wedge \star_g \mbf{A} = \dfrac{1}{2} \int \dd^2x \, \sqrt{-g}~ g^{\mu \nu} \mbf{A}_{\mu} \mbf{A}_{\nu} \;.
	\end{equation}
	\item Evaluate the stress tensor of $S_{\text{aux.}}$ the usual way by varying with respect to the metric
	\begin{equation}
		\mbf{T}_{\mu \nu} := -\dfrac{2}{\sqrt{-g}} \dfrac{\delta S_{\text{aux.}}}{\delta g^{\mu \nu}} \;.
	\end{equation}
	\item  Let the desired stress tensor of the chiral 1-form $A_g$ (i.e. $A_g = \star_g A_g$) be denoted by $T_{\mu \nu}$. Then we can extract $T_{\mu \nu}$ from $\mbf{T}_{\mu \nu}$ as
	\begin{equation}
		T_{\mu \nu} = \mbf{T}_{\mu \nu}\Big\vert_{\mbf{A}=A_g} \quad.
	\end{equation}
\end{enumerate}

In the rest of this section we derive the steps outlined above. The reader uninterested in this derivation can safely skip to the next section once they have taken note of the final expression \cref{eq:T_chiral}.

\subsection{Relating Forms and Pseudoforms}\label{sec:properties}

 Due to the unusual structure of Sen's action, the only term contributing to the stress tensor in an arbitrary background is
\begin{equation}\label{eq:stress_action}
	S = \int \,  \, A \wedge \mc{M}(A) \;.
\end{equation}
Recall that $A$ is an arbitrary self-dual form with respect to the Minkowski metric, even though the theory is defined on a two dimensional manifold $ \mathbb{M} $ with a non-flat metric $g_{\mu \nu}$. $\mc{M}$ is a linear map from the space of self-dual pseudoforms to space of anti-self-dual pseudoforms. In component form, the map acts as $\mc{M}(A)_\nu = \mc{M}_\nu^{\;\;\mu} A_\mu$. The properties of $\mc{M}(\cdot)$ can then be translated into equivalent statements about the tensor $\mc{M}_\nu^{\;\;\mu}$.

Let us now consider a basis of $ 1 $-forms in two dimensions: $\lbrace v_+\,,\,v_-\rbrace$. In fact, in a coordinate basis we have
\begin{equation}\label{eq:v_properties}
	v_+ = \dd x^+ \quad \text{and} \quad v_- = \dd x^- \; \implies \; v_+ \wedge v_- = 2 \, \dd x^1 \wedge \dd x^2.
\end{equation}
It follows that any self-dual $ 1 $-form can be expanded as $A = a_{\mu}(x^+) v_+^{\mu}$, we will abbreviate this often as $A = \mbf{a}\, v_+$, with the boldface serving to remind us it is a component of a $ 1 $-form. We will now try to pin down a precise relation between the basis of chiral forms with respect to $\star_g$ and the basis $v_{\pm}$. 

Let us suppose we have a basis form $\vt_+$ such that $\vt_+ = \star_g \vt_+$. Locally, one can always write down a relation of the form
\begin{equation}\label{eq:basis_change}
	\vt_+^\mu = \mc{T}^{\mu}_{\;\;\;\nu} \, v^\nu_+ + \mc{U}^{\mu}_{\;\;\;\nu} \, v^\nu_-
\end{equation}

Now, due to \cref{eq:Ag_selfdual,eq:A_g}, we know $v_+ - \mc{M}(v_+)$ to be a self-dual 1-form with respect to $\star_g$. Thus we expect it to be linearly related to $\vt_+$. We can write
\begin{align} \label{eq:vg_to_vt}
	\Big[v_+ - \mc{M}(v_+)\Big]^\nu &= \mc{P}^{\nu}_{\;\;\; \mu} \, \vt_+^\mu \nonumber\\
	&=  \mc{P}^{\nu}_{\;\;\; \a} \, \mc{T}^{\a}_{\;\;\;\b} \, v^\b_+ + \mc{P}^{\nu}_{\;\;\; \a}\,\mc{U}^{\a}_{\;\;\;\b} \,v^\b_- \;,
\end{align}
where in the last line we have made use of \cref{eq:basis_change}. Comparing the coefficient of $v_{+}$ on both sides, we get
\begin{align} \label{eq:P_to_T}
	\mc{P}^{\nu}_{\;\;\; \mu}   \mc{T}^{\mu}_{\;\;\; \a} &= \delta^{\nu}_{\;\;\; \a} \nonumber \\
	\implies	\mc{P}^{\nu}_{\;\;\; \mu} &= \left(  \mc{T}^{-1} \right)^{\nu}_{\;\;\; \mu} \;, 
\end{align}
whereas the coefficient of $v_-$ gives (in matrix notation)
\begin{equation}\label{eq:M_to_T}
	\mc{M} = - \mc{T}^{-1}\,\mc{U} \;.
\end{equation}

Now note that we can write
\begin{align}\label{eq:vt_to_v}
	\mc{T}^{-1} \vt_+ &= v_+ +  \mc{T}^{-1}\,\mc{U}\, v_- \nonumber \\
	&= v_+ \;,
\end{align}
where we made use of \cref{eq:M_on_anti,eq:M_to_T}.

It is important to note that $A_g$ is not just equal to $\tfrac{1}{2}(1+ \star_g) A$ as one might naively expect. Rather, we have from the definition in \cref{eq:A_g}
\begin{equation}\label{eq:Ag_to_A}
	A_g = \underbrace{\dfrac{1}{2} \left(A + \star_g A\right)}_{\text{naively expected}} -\, \dfrac{1}{2} \Big(\mc{M}(A) + \star_g \mc{M}(A)\Big) \;.
\end{equation}

A consequence of self-duality condition is that under a variation with respect to the metric, the anti-self-dual part of the variation is constrained to be
\begin{equation}\label{eq:delta_Ag}
	\dfrac{1}{2}\left(1-\star_g\right)\delta_g A_g = \dfrac{1}{2} (\delta_g \star_g) A_g \;.
\end{equation}
On the other hand, the self-dual part of the variation can be set to zero \citep{Gran:2014lia}. The \cref{eq:delta_Ag} will play an important role when we evaluate the stress tensor of the $\ttb$ deformed Lagrangian.

Finally, note that for $A = \mbf{a}\,v_+$, using \cref{eq:vt_to_v} we can express $A_g$ as
\begin{equation}\label{eq:Ag_in_vt}
	A_g = \mc{T}^{-1} \,\mbf{a} \, \vt_+ \;.
\end{equation}

\subsection{Evaluating the stress tensor}

We define the stress tensor the usual way for a QFT coupled to a curved background:
\begin{equation}\label{eq:T}
	T_{\mu \nu} := -\dfrac{2}{\sqrt{-g}} \dfrac{\partial \mc{L}}{\partial g^{\mu \nu}} \;.
\end{equation}
In Sen's action, the only term which will contribute in the action is the term reproduced in \cref{eq:stress_action}. Recall also that $A = \mbf{a}_\mu v_+^\mu$. We can then write the stress-tensor as
\begin{equation}\label{eq:T1}
	T_{\mu \nu} = -\dfrac{2}{\sqrt{-g}}\, \mbf{a}_\mu \, \mbf{a}_\nu \,v_+^{\mu} \wedge \dfrac{\partial \mc{M}}{\partial g^{\mu \nu}}(v_+^{\nu}) \;.
\end{equation}

Since by construction $v_+ - \mc{M}(v_+)$ is self dual with respect to $\star_g$ (recall \cref{eq:Ag_selfdual}), we can vary the self-duality condition with respect to the metric and obtain
\begin{equation}\label{eq:deltaM}
	\left(1-\star_g\right) \delta_g \mc{M}(v_+) = - (\delta_g\star_g) \Big( v_+ - \mc{M}(v_+) \Big) \;.
\end{equation}

Hence, for $\vt_+ = \star_g \vt_+$, we get
\begin{equation}\label{eq:vt_deltaM}
	2 \, \vt_+ \wedge \delta_g \mc{M}(v_+) = - \vt_+ \wedge (\delta_g\star_g) \Big( v_+ - \mc{M}(v_+) \Big) \;.
\end{equation}

Writing the indices explicitly and using \cref{eq:vg_to_vt,eq:P_to_T} to rewrite the right-hand side, we obtain
\begin{equation}\label{eq:step1}
	2 \delta_g \,\mc{M}^{\a}_{\;\;\; \nu} \, \vt_+^{\mu} \wedge v_+^{\nu} = - (\mc{T}^{-1})^{\alpha}_{\;\;\; \b} \,\vt_+^{\mu} \wedge (\delta_g\star_g) \,\vt_+^{\b} \;.
\end{equation}

It follows from \cref{eq:basis_change} that $\vt_+^{\mu} \wedge v_+^{\nu} = \mc{T}^{\mu}_{\;\;\;\a} \, v_+^{\a} \wedge v_+^{\nu}$. Substituting this into the left-hand side of \cref{eq:step1}, we get
\begin{align}\label{eq:final_steps}
	2 \delta_g \,\mc{M}^{\a}_{\;\;\; \nu} \,\mc{T}^{\mu}_{\;\;\;\b}\, v_+^{\b} \wedge v_+^{\nu} &= - (\mc{T}^{-1})^{\alpha}_{\;\;\; \b} \,\vt_+^{\mu} \wedge (\delta_g\star_g) \,\vt_+^{\b} \nonumber\\
	\delta_g \,\mc{M}^{\a}_{\;\;\; \nu}\, v_+^{\s} \wedge v_+^{\nu} &= -\dfrac{1}{2} (\mc{T}^{-1})^{\s}_{\;\;\; \mu} \, (\mc{T}^{-1})^{\alpha}_{\;\;\; \b} \,\vt_+^{\mu} \wedge (\delta_g\star_g) \,\vt_+^{\b} \nonumber \\
	&= -\dfrac{1}{2} \Big( v_+ - \mc{M}(v_+)   \Big)^{\s} \wedge (\delta_g\star_g) \Big( v_+ - \mc{M}(v_+)   \Big)^{\alpha} \;.
\end{align}
Contracting the right-hand side with $\mbf{a}_{\s}$ and $\mbf{a}_{\a}$, the stress tensor simplifies to 
\begin{align}\label{final_T}
	T_{\mu \nu} &= \dfrac{1}{\sqrt{-g}} \Big( A - \mc{M}(A)   \Big) \wedge \dfrac{\partial \, \star_g}{\partial g^{\mu \nu}} \Big( A - \mc{M}(A)   \Big) \nonumber \\
	&= \dfrac{1}{\sqrt{-g}} A_g \wedge \dfrac{\partial \, \star_g}{\partial g^{\mu \nu}} \,A_g   \;.
\end{align}
This stress tensor has an alternative interpretation, one that is better suited for carrying out the explicit computation, which we now explain. Consider the action 
\begin{equation}\label{pseudoaction}
		S_{\text{aux.}} = \dfrac{1}{2} \int_{\mbb{M},~g}   \mbf{A} \wedge \star_g \mbf{A} = \dfrac{1}{2} \int \dd^2x \sqrt{-g} ~ g^{\mu \nu} \mbf{A}_{\mu} \mbf{A}_{\nu} \;.
\end{equation}
where $\mbf{A}$ is an \emph{arbitrary} 1-form field defined on $\mbb{M}$ . The stress tensor corresponding to this action is readily seen to be
\begin{align}
	\mbf{T}_{\mu \nu} &=- \dfrac{1}{\sqrt{-g}} \mbf{A} \wedge \dfrac{\partial \, \star_g}{\partial g^{\mu \nu}} \mbf{A} \; \label{eq:pseduo-stress} \\
					  &=\dfrac{1}{4} g_{\mu \nu} \mbf{A}_{\s} \mbf{A}^{\s} - \dfrac{1}{2} \mbf{A}_{\mu} \mbf{A}_{\nu} \;. \label{eq:explicit-stress}
\end{align}
Comparing \cref{eq:pseduo-stress} with \cref{final_T}, we see our intended stress tensor can be read off as 
\begin{equation}\label{key}
		T_{\mu \nu} = \mbf{T}_{\mu \nu}\Big\vert_{\mbf{A}=A_g} \quad.
\end{equation}
This completes the derivation of the prescription outlined in the beginning of this section.

All that remains is to make use of this prescription to obtain the stress tensor for the chiral boson. Using the explicit expression \cref{eq:explicit-stress} and substituting $\mbf{A} \rightarrow A_g$ we get the desired answer
\begin{equation}\label{eq:T_chiral}
	T_{\mu \nu} =- \dfrac{1}{2} A^g_{\mu} \, A^g_{\nu} \;.
\end{equation}

\section{$ \ttb $ Deformations of Chiral Bosons}
\label{sec:TTbChiralBosons}
We now have all the ingredients required to compute the $\ttb$ deformations of chiral bosons. Of course, to ensure that the deformation does not trivially vanish, we must have both left- \emph{and} right-chiral bosons in our undeformed action. We start with simplest case, when we have one left- and one right-chiral boson, and subsequently generalise this to an arbitrary number of left- and right-chiral fields.

Let us recall some basics. The $ \ttb $ ``flow'' deforms the classical Lagrangian of the seed theory according to the following linear differential equation
\begin{equation}\label{eq:TTbDefinition}
	\frac{\dd \mathcal{L}_\lambda}{\dd \lambda} = \det T_{\mu\nu}^{(\lambda)} = \frac{1}{2} \epsilon^{\mu \nu} \epsilon^{\rho \sigma} T_{\mu \rho}^{(\lambda)} T_{\nu \sigma}^{(\lambda)}\ ,
\end{equation}


and the determinant operator that triggers the flow is defined via point-splitting, which we assume implicitly. We interpret all objects in the above equation as power series in $ \lambda $, the $ \ttb $ coupling constant. This means:
\begin{equation}\label{key}
	\mathcal{L}_\lambda = \mathcal{L}_0 + \lambda \,\mathcal{L}_1 + \lambda^2\, \mathcal{L}_2 + \cdots \ ,
\end{equation}
and
\begin{equation}\label{key}
	T_{\mu\nu}^{(\lambda)} = T_{\mu\nu}^{(0)} + \lambda \,T_{\mu\nu}^{(1)} + \lambda^2 \,T_{\mu\nu}^{(2)} + \cdots \ .
\end{equation}
The goal of the following two sections will be to determine the corrections $ \mathcal{L}_{k\geq 1} $ to the classical Lagrangian, using the methods discussed in the previous sections.

It is natural to expect that the deformed theory will continue to respect Lorentz invariance. Before explicitly computing them, we might ask: what kinds of terms might arise as deformations? It is easy to see that the only non-trivial Lorentz invariant combination of the left- and right-chiral fields one can construct are inner products between them, so on general grounds we expect that
\begin{equation}
\mathcal{L}_k \propto f_k\left(A\cdot B\right) \ ,
\end{equation}
where $ f_k $ are some arbitrary functions that we will now determine.

\subsection{A Pair of Chiral Bosons}
In this section we study a pair of chiral bosons, i.e.~one left- and one right-chiral boson. 
\subsubsection{The General Form of Deformations}
The stress energy tensor for anti-self-dual field $B_\mu$ which takes the same form as self dual form field
\begin{align}
T_{\gamma \delta}^{(a)}&=-\frac{1}{2}B^g_{\gamma} B^g_{\delta}~,
\end{align}
in which case the total stress tensor for a single self-dual and anti-self-dual field is 
\begin{align}\label{eq:T0}
T_{\mu \nu}^{(0)}=-\frac{1}{2} [A^g_{\mu} A^g_{\nu}+B^g_{\mu} B^g_{\nu}]~.
\end{align}
Using \cref{eq:TTbDefinition}, the first deformation of Lagrangian can be computed:
\begin{align}
\mathcal{L}_1 = \frac{1}{\sqrt{-g}}\det \left( T_{\mu \nu}^{(0)} \right)=\frac{\sqrt{-g}}{4} \big(A_{g}\cdot B_{g} \big)^2 \ .
\end{align}

Note that since the $\det T_{\mu \nu} $ is a tensor density of weight $(-2)$, for the Lagrangian to have the correct measure in curved background it must be multiplied by a weight $+1$ object, which is naturally the inverse of $\sqrt{-g}$.\footnote{We are deeply grateful to Dmitri Sorokin for pointing this out to us, as well as correcting an oversight pertaining to this point in an earlier version of the draft.}

Since $A_{g}$ is self-dual, the variation $\delta_g A_{g}$ is necessarily anti-self-dual with respect to $\star_g$. Similarly, we can say that $\delta_g B_{g}$ is necessarily self-dual with respect to $\star_g$. Therefore, we have
\begin{align}
\frac{\partial \mathcal{L }_1}{\partial g^{\mu \nu}}&= \frac{1}{4}\left[\frac{-1}{2\sqrt{-g}}\frac{\partial g}{\partial g^{\mu \nu}} \left( A_{g}\cdot B_{g}\right)^2 + 2 \sqrt{-g} \left( A_{g}\cdot B_{g}\right) A^g_{\alpha }\,B^g_{\beta } \,\frac{\partial g^{\alpha \beta}}{\partial g^{\mu \nu}}\right]~,
\end{align}
where we have taken 
\begin{equation}
\frac{\partial A^g_{\alpha}}{\partial g^{\mu \nu}} \,B_{g}^{\alpha} = 0 \quad \quad \text{and} \quad \quad A_{g}^{\alpha}\, \frac{\partial B^g_{\alpha}}{\partial g^{\mu \nu}} = 0~.
\end{equation}
This implies 
\begin{equation}\label{eq:L1_pair}
\begin{aligned}
\frac{\partial \mathcal{L }_1}{ \partial g^{\mu \nu}}&=-\frac{\sqrt{-g}}{8} \left [\, g_{\mu \nu} \left( A_{g}\cdot B_{g}\right)^2 -  2\left( A_{g}\cdot B_{g}\right) \, A^g_{(\mu }B^g_{\nu)}  \right]  \\
&=\frac{\sqrt{-g}}{8} \ g_{\mu \nu} \left( A_{g}\cdot B_{g}\right)^2\ .
\end{aligned}
\end{equation}
The simplification in \cref{eq:L1_pair} is made possible using the identity \cref{ap:tens}, derived explicitly in Appendix \ref{app:A}.
The first-order deformation of the stress tensor is then calculated as
\begin{align}\label{eq:T1_expr}
T_{\mu \nu}^{(1)} =-\frac{2}{\sqrt{-g}}\frac{\partial \mathcal{L}_1}{\partial g^{\mu \nu}} = -\frac{1}{4} \ g_{\mu \nu} \left( A_{g}\cdot B_{g}\right)^2\ .
\end{align} 
The total stress tensor upto $\mathcal{O}(\lambda)$ is given by $ T_{\mu \nu}^{(\lambda)}= T^{\mu \nu}_{(0)}+\lambda T^{\mu \nu}_{(1)}$. So the determinant of $ T_{\mu \nu}^{(\lambda)}$ becomes
\begin{align}
\det\left(  T_{\mu \nu}^{(\lambda)}\right)= \frac{1}{2}\epsilon ^{\alpha \mu} \epsilon^{\nu \beta} \left(  T_{\mu \nu}^{(0)}+\lambda T_{\mu \nu }^{(1)}\right) \left( T_{\alpha \beta}^{(0)}+\lambda T_{\alpha \beta}^{(1)}\right) \ .
\end{align} 
However to compute the second-order deformation of the Lagrangian, we must pick up $\mathcal{O}(\lambda)$ terms in the determinant. We find that
\begin{align}
\mathcal{L}_2= \frac{1}{4\sqrt{-g}}\epsilon ^{\alpha \mu} \epsilon^{\beta\nu}  T_{\mu \nu}^{(0)} T_{\alpha \beta}^{(1)} =0 \ .   
\end{align} 
Since $\mathcal{L}_2=0$, we see $ T_{\mu \nu}^{(2)} = 0$ as well. In fact, by induction, it can be shown that all even orders of the deformation vanish, i.e.~$\mathcal{L }_{2n} = 0$. They are  calculated from 
\begin{equation}
\sum_{i + j = 2n -1}\epsilon ^{\alpha \mu} \epsilon^{\beta\nu} \  T_{\mu \nu}^{(i)}   T_{\alpha \beta}^{(j)}  =  \epsilon ^{\alpha \mu} \epsilon^{\beta\nu} \  T_{\mu \nu}^{(0)}  T_{\alpha \beta}^{(2n-1)} ~,
\end{equation}
which is identically zero. The odd orders are, however, non-vanishing. To compute $\mathcal{L}_3$ , we pick $\mathcal{O}(\lambda^2)$ terms in $\det\left(  T_{\mu \nu}^{(\lambda)}\right)$, which gives
\begin{equation}
\begin{aligned}
\mathcal{L}_3&= \frac{1}{3 \sqrt{-g}} \det \left(  T_{\mu \nu}^{(1)} \right)  \\
& = -\frac{ \sqrt{-g}}{48}  \left( A_{g}\cdot B_{g}\right)^4 \ .
\end{aligned}
\end{equation} 
and 
\begin{align}
T_{\mu \nu}^{(3)} = \frac{1}{16} \, g_{\mu \nu}\left( A_{g}\cdot B_{g}\right)^4~.
\end{align}
While iterating the algorithm to higher orders, a pattern of deformations emerges: all odd orders in the $ \ttb $ coupling constant $ \lambda $ of the stress tensor can be expressed as $g_{\mu \nu}$ multiplied by even powers of $(A_g \cdot B_g)$, leading us to conclude that
\begin{equation}
\mathcal{L}_{2n-1} \propto \sqrt{-g}  \ \left( A_{g}\cdot B_{g}\right)^{2n} \ .
\end{equation}

This uniform pattern allows us to assume an ansatz. In the following section, we solve \cref{eq:TTbDefinition} to all orders. This will allow us to derive the deformed Lagrangian in closed form.

\subsubsection{The Flow Equation}
Let us define
\begin{equation}
X = \lambda^2 \left( A_{g}\cdot B_{g}\right)^2 \ .
\end{equation}
For the flow equation we assume the ansatz $\mathcal{L}_\lambda =\mathcal{L}_0 + \widehat{\mathcal{L}}_{(\lambda)}$, where
\begin{align}
\widehat{\mathcal{L}}_{(\lambda)}=  \frac{\sqrt{-g}}{\lambda}F\left(X\right) \ ,
\end{align}
where $ F $ is some function we have to solve for. 
We will require that $ F(X) $ satisfy the boundary condition $F(0)=0$, i.e.~that when the deformation is turned off we recover the seed theory. We now write \cref{eq:TTbDefinition} with the above ansatz. The left-hand side of the flow equation is
\begin{equation}
\frac{\dd \mathcal{\widehat{L}}_\lambda}{\dd \lambda} = -\frac{\sqrt{-g}}{\lambda^2} \left[ F(X) - 2F'(X)X\right] \ .
\end{equation}
We also require the stress tensor, which breaks up into contributions from the seed theory and the deformation terms
\begin{equation}
T_{\mu \nu}= T^{(0)}_{\mu \nu} +  T^{(\lambda)}_{\mu \nu} \ ,
\end{equation}
where
\begin{equation}
\begin{aligned}
T^{(\lambda)}_{\mu \nu}  &= - \frac{2}{\sqrt{-g}} \frac{\partial \widehat{\mathcal{L}}_{(\lambda)}}{\partial g^{\mu \nu}} \\
&= g_{\mu \nu} \frac{F(X)}{\lambda} - \frac{F'(X)}{\lambda} \frac{\partial X}{\partial g^{\mu \nu}} \\
& = \frac{g_{\mu \nu} }{\lambda}\left[F(X) - 2F'(X)X\right] \ .
\end{aligned}
\end{equation}
In obtaining the final line of the preceding expression we made use of the identity \cref{ap:tens}, derived explicitly in Appendix \ref{app:A}.

The determinant of the stress tensor of the deformed theory takes the form:
\begin{equation}
\frac{1}{\sqrt{-g}}\det T_{\mu \nu} = \frac{\sqrt{-g}}{4} \frac{X}{\lambda^2} - \frac{\sqrt{-g}}{\lambda^2}\left[F(X) - 2F'(X)X\right]^2 \ .
\end{equation}
We will construct the $\ttb$ flow equation in the flat spacetime for which we set $g=-1$. In this limit the flow equation \cref{eq:TTbDefinition} becomes
\begin{equation}\label{eq:flow}
\left[F(X) - 2F'(X)X\right]^2 - \left[ F(X) - 2F'(X)X\right] - \frac{X}{4}=0
\end{equation}
which, when completing the square, simplifies to 
\begin{equation}
F(X) - 2XF'(X) =\frac{1}{2} \left[ 1-\sqrt{1+X}\right] \ .
\end{equation}
Note that when completing the square, we have chosen the positive branch to ensure consistency with the boundary condition. Now, the above equation can be solved exactly, and we find
\begin{equation}
F(X) = \frac{1}{2}\left[1-\sqrt{1+X} +\sqrt{X}\sinh ^{-1}\left(\sqrt{X}\,\right)+2C \,\sqrt{X} \right]
\end{equation}
and 
\begin{align}
F'(X) = \frac{1}{4 \sqrt{X}}\left(\sinh^{-1}\left(\sqrt{X}\,\right) + 2C   \right)\ .
\end{align}
Requiring that $ F(X) $ has a smooth limit as $ X \rightarrow 0 $ requires that we set $C = 0$. Consequently, the $\ttb$-deformed Lagrangian is
\begin{equation}\label{eq:DeformedSinglePair}
\mathcal{L }_\lambda = \mathcal{L}_0 + \frac{1}{2\lambda}\left[1-\sqrt{1+X}+\sqrt{X}\sinh ^{-1}\left(\sqrt{X}\,\right)\right] \ .
\end{equation}
The first two terms in $ \hat{\mathcal{L}} $ are strongly reminiscent of the Nambu-Goto action in static gauge --- similar looking terms arise when describing the $ \ttb $-deformed free boson Lagrangian. The last term in the Lagrangian, however, appears novel, and suggests that this is the first non-local action written in Sen's formalism.

With a view towards studying the classical dynamics of this theory in the $ \lambda \rightarrow \infty $ limit, we write down the equations of motion for the $\ttb$-deformed theory:
\begin{align}\label{eq:full_eom}
\dd \left( A + \frac{1}{2}\sinh ^{-1}\left(\sqrt{X}\,\right)\, B\right)&=0  \ ,\\
\dd \left( B - \frac{1}{2}\sinh ^{-1}\left(\sqrt{X}\,\right)\, A \right)&=0 \ .
\end{align}
Note that we are only describing here the \emph{physical} sector of the equations of motion. The decoupled free sector, obviously, remains unaffected by the deformation.

\subsection{Infinite Coupling Limit} \label{sec:large_coupling}

With the fully resummed $ \ttb $-deformed Lagrangian in \cref{eq:DeformedSinglePair}, it is only natural to ask, following our earlier investigations \cite{Chakrabarti:2020pxr}, what happens to this theory in the limit of infinite coupling. In our earlier work, it was important that we work in the Hamiltonian formalism --- this was the only way to sensibly take this limit.

In the present context, the Hamiltonian for Sen's formulation, as derived in \cite{Sen:2019qit}, assumed that all other fields with which the self-dual field interacts are ``ordinary'' fields described by more familiar Lagrangians. This is not the case for the $\ttb$-deformed theory at hand, for which both the interacting (self-dual and anti-self-dual) fields are described using Sen's formalism. While we cannot evaluate the Hamiltonian without a minimal extension of the derivation presented in \cite{Sen:2019qit}, we can hope to make some progress at the level of the equations of motion \cref{eq:full_eom}.  Note that the $\ttb$ coupling $\lambda$ has mass dimension $-2$. The dimensionless coupling obtained from $\lambda$ is of the form $m^2\lambda =\tilde{\lambda}$, for some mass scale $m$. The limit $\tilde{\lambda} \rightarrow \infty$ is, of course, well-defined. The $\lambda \rightarrow \infty$ is to be understood as the scenario where one adopts units where the mass scale $m$ is set to unity.

Using the following expansion of $\sinh ^{-1}(\sqrt{X})$ at large $X$
\begin{align}
	\sinh ^{-1}\sqrt{X}=\frac{1}{2}\Big[2 \log 2 +\log X \Big]+ \mathcal{O}\left(\frac{1}{X}\right)~,
\end{align}   
 we can write
\begin{align}
	\lim _{\lambda \rightarrow \infty} \sinh ^{-1}\sqrt{X}= \ln \big(\lambda\, A \cdot B \big)~ = \ln\, \lambda + \ln \big(A \cdot B\big)~.
\end{align} 
So in the $\lambda \rightarrow \infty $ limit, dropping all terms which are $\lambda$ independent or subleading, the equation of motions of the $\ttb$-deformed theory simplifies to
\begin{align}
	\text{d}\left[ \log \lambda ~B\right]=0 \Rightarrow \text{d}B=0~,\\
	\text{d}\left[\log \lambda ~A\right]=0 \Rightarrow  \text{d}A=0~.
\end{align}

These equations are nothing but the equations of the seed theory on a flat background. Therefore, we observe chiral decoupling for a $\ttb$-deformed theory of chiral bosons in the limit of infinite coupling.

\subsection{Arbitrary Number of Chiral Bosons}
A similar calculation can be carried out for an arbitrary (but non-zero) number of self-dual and anti-self-dual bosons. The stress tensor of the undeformed Lagrangian can be written as
\begin{align}
T_{\mu \nu}^{(0)} =-\frac{1}{2} \sum_{i=1}^{N_c} A_{(g)\mu}^i \, A_{(g)\nu}^i \, -\frac{1}{2} \sum_{j=1}^{N_a} B_{(g)\mu}^j B_{(g)\nu}^j~,
\end{align}
where $A_{(g)}^i$ and $B_{(g)}^j$ are left- and right-chiral bosons, and $N_c$ and $N_a$ denote the number of such bosons respectively.
As before, the first deformation is given by
\begin{align}
\mathcal{L}_1 = \frac{1}{\sqrt{-g}}\det\left(T_{\mu \nu}^{(0)}\right)= \frac{\sqrt{-g}}{4} \sum_{i = 1}^{N_c}\sum_{j= 1}^{N_a}\Big(A_{(g)}^i \cdot B_{(g)}^j \Big)^2~,
\end{align}
since $A^i_{(g)} \cdot A^j_{(g)} = 0$ and $B^i_{(g)} \cdot B^j_{(g)} = 0$.

The next order of the stress tensor is calculated by varying $\mathcal{L}_1$ with respect to the metric, as was done previously. This gives us
\begin{align}
T_{\mu \nu}^{(1)} = -\frac{1}{4} \ g_{\mu \nu}  \left [\sum_{i = 1}^{N_c}\sum_{j= 1}^{N_a} \left( A_{(g)}^i\cdot B_{(g)}^j\right)^2 \right]  ~,
\end{align}
The $ \mathcal{O}(\lambda^2) $ deformation $\mathcal{L }_2$ is calculated from the $\mathcal{O}(\lambda)$ term as before, and is given in terms of the determinant of $T_{\mu \nu}^{(\lambda)}=T_{\mu \nu}^{(0)}+\lambda T_{\mu \nu}^{(1)}$ as
\begin{equation}
\mathcal{L}_2 = \frac{1}{4\sqrt{-g}}\epsilon^{\alpha \alpha'} \epsilon^{\beta \beta'}T_{\alpha\beta}^{(0)} T_{\alpha' \beta'}^{(1)}~. \qquad 
\end{equation}
When explicitly evaluated, we see that every term contains terms of the form $(A^i_{(g)} \cdot A_{(g)}^j)$ or $(B_{(g)}^i \cdot B_{(g)}^j)$ which vanish identically, implying $\mathcal{L }_2 = 0$. 

In fact, from the structure of the deformations, it can be shown that all even orders of the deformation vanish, i.e.~$\mathcal{L }_{2n} = 0$ for all $n \geq 1$ for the same reason: they are calculated from the terms of the form
\begin{equation}
\epsilon^{\alpha \alpha'} \epsilon^{\beta \beta'}T_{\alpha\beta}^{(i)} T_{\alpha' \beta'}^{(j)}
\end{equation}
with $ i+j = 2n-1 $. Since $ T_{\alpha\beta}^{(\text{even})} = 0 $ the only non-zero term among these is the one with $ (i,j) = (0,2n-1) $. When expanded, this terms always come with the contractions $(A^i_{(g)} \cdot A^j_{(g)})$ or $(B^i_{(g)} \cdot B^j_{(g)})$. Since these contractions are identically zero, the conclusion that all even order deformations vanishes follows. 

However, the procedure we have outlined suggests that odd orders are non-vanishing. For example, $\mathcal{O}(\lambda^3)$ term of the deformed Lagrangian is given by 
\begin{equation}
\begin{aligned}
\mathcal{L}_3&= \frac{1}{3 \sqrt{-g}} \det \left(  T_{\mu \nu}^{(1)} \right) \\
& = -\frac{ \sqrt{-g}}{48}  \left [\sum_{i = 1}^{N_c}\sum_{j= 1}^{N_a} \left( A_{(g)}^i\cdot B_{(g)}^j\right)^2 \right]^2 \ .
\end{aligned}
\end{equation}
Observe that, 
\begin{equation}
\begin{aligned}
\frac{\partial}{\partial g^{\mu \nu}}\left [\sum_{i = 1}^{N_c}\sum_{j= 1}^{N_a} \left( A_{(g)}^i\cdot B_{(g)}^j\right)^2 \right] &= \left[\sum_{i = 1}^{N_c}\sum_{j= 1}^{N_a} \left( A_{(g)}^i\cdot B_{(g)}^j\right) \left( A_{(g)(\mu}^iB_{(g)\nu)}^j\right)\right] \\
&= g_{\mu \nu} \left[\sum_{i = 1}^{N_c}\sum_{j= 1}^{N_a} \left( A_{(g)}^i\cdot B_{(g)}^j\right)^2 \right] \ ,
\end{aligned}
\end{equation}
where we have used the identity \cref{ap:tens} given in Appendix \ref{app:A}. Therefore, by using the chain rule, one can see that
\begin{equation}\label{eq:deriv_of_arbit}
\begin{aligned}
\frac{\partial}{\partial g^{\mu \nu}}\left [\sum_{i = 1}^{N_c}\sum_{j= 1}^{N_a} \left( A_{(g)}^i\cdot B_{(g)}^j\right)^2 \right]^n= n\, g_{\mu \nu} \left[\sum_{i = 1}^{N_c}\sum_{j= 1}^{N_a}\left( A_{(g)}^i\cdot B_{(g)}^j\right)^2 \right]^n \ .
\end{aligned}
\end{equation}
The term in the square brackets is analogous to $\left(A_g \cdot B_g\right)^{2}$ from the case of a pair of chiral bosons. Using \cref{eq:deriv_of_arbit} we can easily calculate $T^{(3)}_{\mu \nu}$ as
\begin{align}
T_{\mu \nu}^{(3)} = \frac{1}{16} \, g_{\mu \nu}\left [\sum_{i = 1}^{N_c}\sum_{j= 1}^{N_a} \left( A_{(g)}^i\cdot B_{(g)}^j\right)^2 \right]^2 ~.
\end{align}
From this discussion, it is clear that a new variable analogous to $X$ appropriate to the case of arbitrary numbers of left- and right-chiral bosons can be defined as
\begin{align}
X=\lambda ^2 \left [\sum_{i = 1}^{N_c}\sum_{j= 1}^{N_a} \left( A_{(g)}^i\cdot B_{(g)}^j\right)^2 \right] \ .
\end{align}
With this $ X $, the closed form expression for the $ \ttb $-deformed Lagrangian is derived in a manner similar to the previous section. Thus, the $ \ttb $-deformed Lagrangian corresponding to the case of an arbitrary number of left- and right-chiral bosons is given by \cref{eq:flow}.

Before proceeding, let us comment on the question of curved spacetime, since our choice of formalism is naturally developed for any Lorentzian $ 2 $-manifold $ \mathbb{M} $ with an arbitrary metric $ g $. There is at present no consensus regarding status of $\ttb$ deformations in an arbitrary curved background, although some progress has been made for spacetimes with constant curvature in \cite{Jiang:2019tcq}. It was found that the paradigmatic properties of the $ \ttb $ deformation such as Zamolodchikov's factorisation formula no longer continue to hold, and one finds curvature-induced corrections.

To deform a two-dimensional quantum theory on a curved background, we must first define the operator $\ttb$ as usual, via point-splitting. For curved backgrounds this involves talking about quantities defined at different spacetime points --- this is, in essence, the source of the difficulty. It is not our goal to propose a general resolution to this problem. Rather, we observe that for the specific theories under consideration in this paper, Sen's formalism enables us to express the theory \emph{at all points} using pseudoforms. The pseudoforms themselves are completely oblivious to the background metric and therefore allow for the comparison of fields at different spacetime points. It seems to us that Sen's formalism is well positioned to investigate $ \ttb $ deformations on curved spacetimes in greater detail. We leave this study for the future.

It is well known that a theory with a chiral anomaly, on being coupled to gravity produces a theory with a gravitational anomaly. While at the classical level, these anomalies play no role, a theory with an arbitrary number of left- and right-chiral bosons coupled to background metric is only consistent quantum mechanically when the original theory has chiral symmetry, i.e.~when $N_c = N_a$. Since we focus solely on flat spacetime, these considerations have no bearing on our analysis.

Before concluding, let us remark that a similar analysis following \cref{sec:large_coupling} can be done for the case of arbitrary left and right chiral bosons as well. The conclusion, as can be easily checked, remains the same. In the general case also, the equations of motion exhibit chiral decoupling in the limit of infinite coupling.

\section{Fermions and Bosonisation}
\label{sec:Bosonisation}
In this section we will solely work in flat space. In this case the pseudoform coincides with the physical form that is self-dual with respect to $\star$, the flat space Hodge dual.

\subsection{Fermions}

Let us recall the $ \ttb $-deformed free fermion Lagrangian \cite{Coleman:2019dvf}, which is
\begin{equation}\label{eq:fermion_action}
	\begin{aligned}
		\mathcal{L}_{\lambda} &= \mathcal{L}_0 + \mathcal{L}_{\text{def}.} \\
		&=\psi_{-}^{\dagger} \partial \psi_{-}+\psi_{+}^{\dagger} \bar{\partial} \psi_{+}+\lambda\left[\left(\psi_{-}^{\dagger} \partial \psi_{-}\right)\left(\psi_{+}^{\dagger} \bar{\partial} \psi_{+}\right)-\left(\psi_{-}^{\dagger} \bar{\partial} \psi_{-}\right)\left(\psi_{+}^{\dagger} \partial \psi_{+}\right)\right] \ ,
	\end{aligned}
\end{equation}
where $ \mathcal{L}_{\text{def}.} $ refers to the term linear in $ \lambda $. It was observed that the deformations truncate at linear order in the $ \ttb $ coupling. In fact, something more drastic is true: the first-order (and consequently all subsequent orders) deformation of the stress tensor vanishes identically! That is,
\begin{equation}\label{key}
	 T_{\mu\nu}^{(1)} = 0 \ .
\end{equation}
To summarise, we have a seed conformal field theory and an irrelevant deformation thereof, both of which have the same (traceless, of course) stress tensor, the same degrees of freedom, and the same central charge. To the best of our knowledge, this has not been remarked on in the literature. 

That the stress tensors of the seed theory and the deformed theory coincide, despite the fact that the deformed theory contains an interaction term, is highly unusual. The interaction term, it would seem, has no impact on the energetics of the system. 

A possible resolution to this puzzle that immediately comes to mind is that the free and deformed Lagrangians are related by a nonlinear field redefinition. If this is the case, one diagnostic in the absence of the explicit field redefinition would be to compute the S-matrices of the deformed theory, which should be trivial.

Another possibility that might explain this observation is that the deformation term is purely topological by nature. In fact, as we will now explain, this indeed is the correct interpretation. Consider the following $ 1 $-forms
\begin{equation}\label{S+-}
	S^+_{\mu} := \lambda \, \psi_{+}^{\dagger} \partial_{\mu} \psi_{+} \quad \text{and} \quad S^-_{\mu} := \lambda \, \psi_{-}^{\dagger} \partial_{\mu} \psi_{-} ~.
\end{equation}
The prefactor $\lambda$ ensures that $S^{\pm}$ has the mass dimension expected of a $ 1 $-form. Then one can readily check that the deformation term of \cref{eq:fermion_action} can be expressed as 
\begin{equation}\label{eq:topology}
	\mathcal{L}_{\text{def}.} = \frac{1}{\lambda}\int S^+ \wedge S^- = \frac{1}{\lambda}\int d^2x \, \e^{\mu \nu}\, S^+_{\mu}\, S^-_{\nu} ~.
\end{equation}

Clearly, this term is purely topological and naturally does not contribute to the stress tensor. 
 This line of investigation might be able to explain a puzzle we raised in \cite{Chakrabarti:2020pxr}: how is it possible for strongly interacting chiral fermions to, in the infinite coupling limit, decouple? Once we take the infinite coupling limit such that $S^{\pm}$ is held fixed, we see that the topological term becomes insignificant, leaving us once again with the free fermion Lagrangian. 
 
 The immutable nature of stress tensor also resolves a puzzle raised in \cite{Coleman:2019dvf}: why does the deformation of multiple free fermions truncate at linear order? The genesis of the puzzle can be traced to the argument that for a single pair of free fermion the truncation of linear order necessarily follows due to the lack of non-trivial higher power of bilinears owing to the grassmann odd nature of $\psi_{\pm}$. Such non-trivial higher power terms, of course , exists once we introduce multiple numbers of left and right chiral fermions. As we argued in this section the truncation at linear order follows due to the fact that the deformed stress tensor is \emph{identical} to the undeformed stress tensor. Therefore any subsequent deformations will generate the same deformation terms and can be re-absorbed in a redefinition of the coupling $\lambda$. This explanation immediately extends to the case of multiple fermions hence resolving the puzzle discussed above.
 
 Another observation regarding the deformation term \cref{eq:topology}, is that it closely resembles the fermionic Wess-Zumino term in the superstring worldsheet action as described by Green-Schwarz formalism \cite{Green:1983wt}. Therefore, the action for deformed fermions in this sense is indeed a fermionic analogue of the Nambu-Goto action.

\subsection{A Bosonisation Map}

Let us consider the deformation of the chiral boson theory once again, but this time we will truncate the deformation to linear order in the $ \ttb $ coupling constant. The resulting Lagrangian is easily seen to be
\begin{equation}
	\mathcal{L}_{\lambda}=\mathcal{L}_{0} + \frac{\lambda}{4}\left(A \cdot B\right)^{2} \ .
\end{equation}
It is tempting to conjecture, based on the observation that both theories have free chiral degrees of freedom as seed theories, that after $ \ttb $ deformation at linear order, there might exist a bosonisation duality relating the two deformed theories. Indeed, their respective finite-volume spectra are deformed in a universal manner, dictated by the inviscid Burgers equation.\footnote{We thank Shouvik Datta for emphasising this point.} The fact that for both theories the deformation at linear order continues naturally generalises to an arbitrary number of left- and right-chiral fields further strengthens the case for a possible bosonisation relation. However, as we have already shown, on the bosonic side the deformation leads to an interacting theory, whereas on the fermionic side, the deformation does not appear to alter the classical dynamics, as evinced by the stress tensor remaining unchanged. In this section we speculate on the possibility of an interplay between $ \ttb $ deformations and bosonisation, however far-fetched it may seem, and find that there is some insight to be gained even with the naive comparison. 

Let us focus on the deformation terms. On-shell, the deformation on the fermion side is
\begin{equation}\label{key}
	\lambda \left(\psi_{-}^{\dagger} \bar{\partial} \psi_{-}\right)\left(\psi_{+}^{\dagger} \partial \psi_{+}\right) \ ,
\end{equation}
whereas on the boson side we have
\begin{equation}\label{key}
	\frac{\lambda}{4}\left(A \cdot B\right)^{2} \ .
\end{equation}
Let us write the latter in terms of components, for which we use $ A_\mu = (-\mathfrak{a},\mathfrak{a}) $ and $ B_\mu = (\mathfrak{b},\mathfrak{b}) $ as before. We then find the bosonic interaction term is 
\begin{equation}\label{key}
	\lambda \, \mathfrak{a}^2 \, \mathfrak{b}^2 \ .
\end{equation}
If we were to naively guess a map, then based on considerations of chirality and dimension, we might say
\begin{equation}\label{eq:ConjecturedBosonisationMap}
	\begin{aligned}
		\mathfrak{a}^2 &\longleftrightarrow \psi_{+}^{\dagger} \partial \psi_{+} \ , \\
		\mathfrak{b}^2 &\longleftrightarrow \psi_{-}^{\dagger} \bar{\partial} \psi_{-} \ .
	\end{aligned}
\end{equation}
While these formula do not look like the usual bosonisation correspondence --- between chiral fermions and vertex operators of chiral bosons --- we should keep in mind that $ \mathfrak{a} $ and $ \mathfrak{b} $ are really components of the field strengths. 

If we postulate that $ \mathfrak{a} \sim \partial_+\varphi $ and $ \mathfrak{b} \sim \partial_- \varphi $ on-shell, one can readily check the map proposed above in \cref{eq:ConjecturedBosonisationMap} is related to the more familiar bosonisation formulas. However, we take the perspective that the identification of $\mathfrak{a} \sim \partial_+\varphi $ etc.~is antithetical to the spirit of Sen's formalism, not to mention severely limiting any attempt to generalise this to the corresponding quantum theory. The above map raises an intriguing question: what do bosonisation formulas look like in Sen's formalism? The way forward will be to work entirely within the purview of Sen's formalism on the bosonic side of the map. On the fermionic side, the astute reader may have already noted, the bilinears appearing on RHS of \cref{eq:ConjecturedBosonisationMap} are nothing but self-dual part of the one-form $S^+$ and the anti-self-dual part of the one-form $S^-$ respectively. Another way forward might be to re-write the fermionic free terms using Sen's formalism in terms of the (anti) self-dual parts of the 1-forms $S^{\pm}$. 

We emphasise that this is merely suggestive of the possibility that $ \ttb $ deformations may have an interesting interplay with bosonisation. More adventurously, one can be tempted to conjecture that $\ttb$-deformations ``commute" with bosonisation, i.e.~that bosonisation duality persists under irrelevant deformations. Much more work would be required to establish that this is in fact the case. For the case at hand the bosonisation proposed is between a seemingly free theory (of chiral fermions) and an interacting theory (of chiral bosons), and is therefore not expected to be entirely accurate. Nonetheless, it is natural to wonder if a similar bosonisation map will emerge for the cases where $\ttb$ deformation is non-trivial on both sides. For this question, it might be helpful to start with theories of chiral fermions which are not free but subjected to some potential. 

Indeed, the question of bosonisation can already be asked at the level of the corresponding free theories itself, where due to the non-standard features of Sen's formalism, this already is an interesting question. We expect that the relations proposed in this manuscript might lead us to the correct map. We hope to return to these questions --- in particular, the construction of chiral boson vertex operators in Sen's formalism --- in the future.

\section{Discussion}
\label{sec:Discussion}

In this paper, we have studied the $ \ttb $-deformed Lagrangian corresponding to a theory with an arbitrary number of left- and right-chiral bosons. We have established that the deformations persists at all orders in the $ \ttb $ coupling and we derived the deformed Lagrangian in closed form by suitably constructing and exactly solving the flow equation. The choice to work with Sen's formalism allowed us to present the first instance of a $ \ttb $ deformation of a theory with a non-standard action. 

We also argued that at the limit $\lambda \rightarrow \infty$, the equations of motion of deformed chiral bosons once again exhibits chiral decoupling. This presents a very nice completion of the questions raised by the results of \cite{Chakrabarti:2020pxr} and reviewed in the introduction. Schematically, we have established the following for classical theories of free bosons and their $\ttb$-deformations:

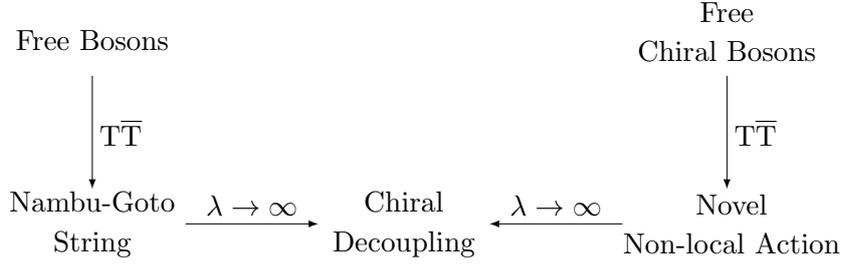
\ctikzfig{diagram}

The limit $\lambda \rightarrow \infty$ was interpreted as ultra-relativistic (also called Carollian) limit in \cite{Blair:2020ops}. There, it was further argued that the chiral decoupling seen in this limit is a natural expectation based on the classification given in \cite{Morand:2017fnv}. Since the non-local action obtained in \cref{eq:DeformedSinglePair} also appears to exhibit chiral decoupling in this limit, this raises the interesting possibility of it being the gauge-fixed form of some kind of worldsheet theory. In any case, this is the first non-local action presented in Sen's formalism, and it is quite interesting on its own merit. We hope to return to a detailed study of this action and its quantization in the future.

For the case of $ \ttb $-deformed free fermions, our analyses indicate that a resolution to a problem we noted in \cite{Chakrabarti:2020pxr} --- why does the infinite coupling limit of the $ \ttb $-deformed free fermion yield free chiral fermions? --- is within reach. In this case, the vanishing of corrections to the stress tensor suggest that the ``interacting'' chiral fermions were really always free. This result immediately explains the observed chiral decoupling of free fermions. Schematically,
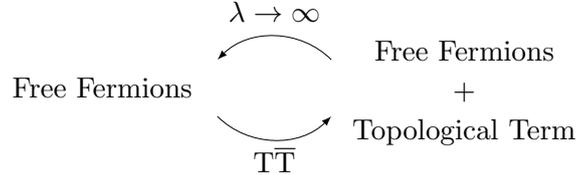
\ctikzfig{Fermion_4}

We also initiated a study on possible interplay between bosonisation and $\ttb$ deformations. In \S\ref{sec:Bosonisation}, we looked at the deformation terms (at linear order) on-shell and proposed a possible bosonisation relation. Our findings imply that if we can find the bosonisation map between the free theories of chiral fermions and bosons, then we can possibly extend such a bosonisation map to the full quantum theory. We also commented on the more tantalising possibility, that $\ttb$-deformations might ``commute'' with bosonisation. However, to shore this correspondence up would require much more work. Firstly, we need to consider seed theory for fermions whose $\ttb$ deformations indeed deforms the dynamics of the theory. In particular one must compute and compare deformed correlation functions. As we have discussed, the standard bosonisation map sends chiral fermions to vertex operators of chiral bosons. An understanding of how bosonisation might manifest in Sen's formalism requires that we first understand how to construct vertex operators in this formalism. It would be an interesting exercise to then check that paradigmatic examples of this duality (for example, the duality between the sine Gordon theory and the massive Thirring model) can be reproduced. We hope to return to these questions in the future.

\acknowledgments{We are grateful to Sujay Ashok, Soumangsu Chakraborty, Shouvik Datta, Subham Dutta Chowdhury, and Ronak Soni for helpful discussions. We especially thank Dmitri Sorokin for his comments on a previous version of this draft. DG acknowledges the hospitality of IMSc, Chennai during the initial stages of this work. MR acknowledges support from the Infosys Endowment for Research into the Quantum Structure of Spacetime.}

\appendix

\section{A Useful Identity}
\label{app:A}
In this appendix we derive an useful identity which proves to be very useful in simplifying the stress tensor for chiral bosons. More specifically, we will show
\begin{align}\label{ap:tens}
	g_{\mu\nu} (A_g \cdot B_g) = A^g_\mu B^g_\nu + A^g_\nu B^g_\mu~,
\end{align}
by explicitly evaluating the components on both sides. To avoid clutter, we drop the subscript $g$ in the rest of this appendix. 

The left-hand side of \cref{ap:tens} when $(\mu,\nu) = (0, 0)$ can be written as
\begin{align}\label{ap:dual}
	g_{00} (A_0 B^0 + A_1 B^1)  =   g_{00} (A_0 B^0 + A^0 B_0)~,
\end{align}
where we have used 
\begin{equation}
	\epsilon^{\mu\nu} A_\nu = \sqrt{-g} A^\mu \quad \text{and} \quad \epsilon^{\mu\nu} B_\nu = -\sqrt{-g} B^\mu \ ,
\end{equation}
to rewrite the second term. This equation can be further simplified as
\begin{align*}
	&\quad \:\: A_0  \left(g_{00}B^0\right) + \left(g_{00}A^0\right) B_0 \\
	&= A_0  \left(B_0 - g_{01}B^1\right) + \left(A_0-g_{01}A^1\right) B_0\\
	&= 2 A_0 B_0 - g_{01} \left(A_0B^1 + A^1B_0 \right)\\
	&= 2 A_0 B_0 ~, \numberthis
\end{align*}
which is the R.H.S of \eqref{ap:tens}. In the second last line, we make use of \eqref{ap:dual}.

Similarly, for $(\mu,\nu) = (1, 1)$,
\begin{align*}\label{ap:dual}
	g_{11} \left( A_0 B^0 + A_1 B^1 \right)  &=   g_{11} (A^1 B_1 + A_1 B^1) \\
	& =  \left(A_1-g_{01}A^0\right) B_1 + A_1  \left(B_1 - g_{01}B^0\right)\\
	&= 2 A_1 B_1 - g_{01} \left(A^0B_1 + A_1B^0 \right)\\
	&= 2 A_1 B_1 ~,\numberthis
\end{align*}
and for $(\mu,\nu) = (1, 0)$, 
\begin{align*}\label{ap:dual}
	g_{10} \left(A_0 B^0 + A_1 B^1\right)  &=   g_{10} \left(A_0 B^0 + A^0 B_0 \right) \\
	& =  A_0\left(B_1-g_{11}B^1\right) +  \left(A_1 - g_{11}A^1\right) B_0\\
	&=  A_0 B_1 + A_1 B_0 - g_{11}\left(A_0B^1 + A^1B_0\right)\\
	&= A_0 B_1 + A_1 B_0 ~.\numberthis
\end{align*} 
This establishes \cref{ap:tens}.

\bibliographystyle{JHEP}
\bibliography{Refs}
\end{document}

%% file: Flowchart.tikzstyles
\tikzstyle{text block}=[fill=none, draw=none, text width=4cm, align=center, tikzit shape=rectangle, tikzit draw=black, tikzit fill=white]

\tikzstyle{arrow}=[draw=black, fill=none, -latex, line width = 0.1mm]